\documentclass{article}

\usepackage[T1]{fontenc}

\usepackage{float}
\usepackage{hyperref}
\usepackage{graphicx}
\graphicspath{{images/}{../images/}}

\usepackage[ruled, linesnumbered, vlined, noend]{algorithm2e}

\usepackage{xcolor}
\usepackage{amssymb, amsthm, amsmath}

\usepackage{nicematrix}
\usepackage{tikz,pgfplots}
\usetikzlibrary{arrows.meta, positioning, calc}
\pgfplotsset{compat=newest}

\usepackage{authblk}

\DeclareMathOperator{\SPEP}{SPEP}
\DeclareMathOperator{\poly}{poly}

\newcommand{\F}{\mathbb{F}}
\newcommand{\Z}{\mathbb{Z}}
\newcommand{\calC}{\mathcal{C}}

\newtheorem{theorem}{Theorem}
\newtheorem{lemma}[theorem]{Lemma}
\newtheorem{prop}[theorem]{Proposition}
\newtheorem{defi}[theorem]{Definition}
\newtheorem{rem}[theorem]{Remark}
\newtheorem{problem}{Problem}

\author[1]{Anna-Lena Horlemann}
\author[2]{Abhinaba Mazumder}
\author[3]{Michael Schaller}
\author[4]{Violetta Weger}

\affil[1]{University of St.Gallen, Switzerland}
\affil[2,3]{University of Zurich, Switzerland}
\affil[4]{Technical University of Munich, Germany}

{
    \makeatletter
    \renewcommand\AB@affilsepx{: \protect\Affilfont}
    \makeatother

    \affil[ ]{Email}

    \makeatletter
    \renewcommand\AB@affilsepx{, \protect\Affilfont}
    \makeatother
    \affil[1]{anna-lena.horlemann@unisg.ch}
    \affil[2]{abhinaba.mazumder@math.uzh.ch}
    \affil[3]{michael.schaller@math.uzh.ch}
    \affil[4]{violetta.weger@tum.de}
}

\title{A Survey on Code Equivalence: The State-of-the-Art and Open Questions}

\makeindex

\begin{document}

\maketitle

\begin{abstract}
In this work, we provide a comprehensive survey of the code equivalence problem and its variants. 
We explain the existing results, highlighting the relationships between different problem formulations, algorithmic techniques, and hardness assumptions. 
In addition, we systematically review known attacks, identify the parameter regimes in which they are effective, and discuss their limitations. 
Lastly, we outline several open problems and research directions, with the aim of clarifying the current landscape and guiding future work toward a deeper understanding of the hardness of code equivalence.
\end{abstract}

\section{Introduction}

The code equivalence problem asks whether two linear codes are equivalent under a Hamming-metric isometry. 
Two prominent variants are commonly studied: the Permutation Equivalence Problem (PEP), where the isometry is restricted to  permutations, and the Linear Equivalence Problem (LEP), where monomial transformations  are allowed.

In recent years, these problems have attracted renewed attention due to their role as hardness assumptions in post-quantum cryptography, where they underpin several cryptographic constructions~\cite{biasse2020less, barenghi2022advanced, battagliola2025vole, hanzlik2025tanuki, albrecht2025hollow, baldi2025speck}. 
At the same time, a wide range of algorithmic techniques has been developed to attack them, drawing from diverse areas such as combinatorics, algebra, and graph theory. 
These include approaches based on low-weight codewords~\cite{beullens2020not, budroni2025two}, hull properties~\cite{SSA}, graph isomorphism reductions~\cite{magali}, canonical forms~\cite{chou2025linear, nowakowski2025improved}, algebraic methods~\cite{saeed}, and Schur product techniques~\cite{square, battagliola2026powerpowercodesnew}.
A good reference for code equivalence is the thesis of Saeed \cite{saeed}, where several concepts, such as hull, puncturing, and guessing techniques are explained.

Despite this substantial amount of work, our current understanding of the hardness of code equivalence remains incomplete. 
Different techniques apply in different parameter regimes, often with little overlap, and the resulting picture is far from unified. 
For instance, several families of weak instances for PEP are known, leading to efficient (often average-case polynomial-time) algorithms~\cite{SSA, magali, square}. 
In contrast, for LEP the situation appears  different: apart from specific small-field cases~\cite{sendrier2013hardness}, and small dimension cases \cite{battagliola2026powerpowercodesnew}, no broad class of weak instances is known, and generic instances over larger fields are widely conjectured to be hard.
This disparity, together with the diversity of existing techniques, raises several fundamental questions. 
Which variants of code equivalence should be considered secure for cryptographic purposes? 
How do the different algorithmic approaches compare across parameter regimes? 
To what extent can hardness results or attacks be transferred between PEP, LEP, and other related problems? 
At present, answers to these questions are scattered across the literature, making it difficult to obtain a clear and coherent view of the field.

In this survey we address these questions by systematically reviewing existing results and presenting a unified picture of the different variants of code equivalence, their relationships, and their complexity.
We conclude with a discussion of several relatively unexplored approaches and a presentation of open problems and questions we have encountered.

\section{Preliminaries}\label{sec:prelim}

We denote by $\mathbb{F}_q$ the finite field with $q$ elements, where $q$ is a prime power and denote by $\mathbb{F}_q^\star$ its multiplicative group, i.e., $\mathbb{F}_q \setminus \{0\}$. 
The identity matrix of size $k$ is denoted by  $\text{Id}_k$. By $\mathrm{diag}(d_1,\dots,d_n)$ we denote the diagonal matrix with diagonal entries $d_1,\dots, d_n$.
Sets are denoted by upper case letters and for a set $S \subseteq I$, we denote by $| S |$ its cardinality and by $S^C$ its complement, i.e., $I \setminus S$.
By $\text{GL}_n(\mathbb{F}_q)$ we denote the $n \times n$ invertible matrices over $\mathbb{F}_q.$
Finally, we denote by $S_n$ the symmetric group on $n$ elements, i.e., the group of permutations of $\{1, \ldots, n\}.$ By abuse of notation we will also say that an $n \times n$ permutation matrix corresponding to the permutation $\sigma \in S_n$ is in $S_n.$
 For a vector $x \in \mathbb{F}_q^n$ and a set $S \subset \{1, \ldots,n\}$ we denote by  $x_S$ the vector consisting of the entries of $x$ indexed by $S.$ Similarly, for a matrix $A \in \mathbb{F}_q^{m \times n}$ and a subset $S \subseteq \{1, \ldots, n\}$ we denote by $A_S$ the matrix consisting of the columns of $A$ indexed by $S.$
 Moreover, we denote by 
$\langle x, y \rangle := \sum_{i=1}^n x_iy_i$ the  standard inner product.
We write $\operatorname{poly}(x)$ for a function that has polynomial growth in $x$\footnote{where we allow $x$ to be a vector of several variables}.

\begin{defi} 
 Let $1 \leq k \leq n$ be integers. An $[n,k]_q$ \emph{linear code} $\mathcal{C}$ is a $k$-dimensional linear subspace of $\mathbb{F}_q^n$. 
The parameter $n$  is called the \emph{length} of the code,  the elements in the code are called \emph{codewords} and $R=k/n$ is called the \emph{rate} of the code. 
A matrix $G \in \mathbb{F}_q^{k \times n}$ is called a \emph{generator matrix} of an $[n,k]_q$ code $\mathcal{C}$ if
$\mathcal{C} = \left\{ x G \mid x \in \mathbb{F}_q^k\right\}.$
\end{defi}

We will often write $\langle G \rangle$ to denote the code generated by $G.$ Similarly, 
for $x_1, \ldots, x_k \in \mathbb{F}_q^n$, $\langle x_1, \ldots, x_k \rangle$ denotes the subspace generated by these vectors.
An $[n,k]_q$ linear code $\mathcal{C}$ has \emph{information set}  $I \subset \{1, \ldots, n\}$ of size $k$, if $| \mathcal{C}|= | \mathcal{C}_I|,$ for $\mathcal{C}_I=\{c_I \mid c \in \mathcal{C}\}.$
As a consequence, $G_I$ is an invertible $k\times k$ matrix.
For any generator matrix $G \in \mathbb{F}_q^{k \times n}$ there exist  some $n \times n$ permutation matrix $ P$ and some invertible matrix $ U \in \mathbb{F}_q^{k \times k}$ that bring $G$ in  \emph{systematic form}, i.e., 
$ UG P =  \begin{pmatrix}
\text{Id}_{k} &  A
\end{pmatrix}, $ where $ A \in \mathbb{F}_q^{k \times (n-k)}$.

\begin{defi} 
Let  $k \leq n$ be positive integers and let  $\mathcal{C}$ be an $[n,k]_q$ linear  code. The \emph{dual code}  $\mathcal{C}^\perp$ is  an $[n,n-k]_q$ linear code, defined as 
$$\mathcal{C}^\perp := \{ x \in \mathbb{F}_q^n \mid \langle x, y \rangle = 0 \ \forall \ y \in \mathcal{C} \}.$$
A matrix $H \in \mathbb{F}_q^{(n-k) \times n}$ is called a \emph{parity-check matrix} of $\mathcal{C}$, if
it is a generator matrix of $\mathcal{C}^\perp$.
\end{defi}
 
\begin{defi} 
Let $\mathcal{C} \subseteq \mathbb{F}_q^n$ be a linear code. Then the   \emph{hull} of $\mathcal{C}$ is defined as $\mathcal{H}(\mathcal{C}) := \mathcal{C} \cap \mathcal{C}^\perp.$
\end{defi}

Note that the dimension of $\mathcal{H}(\mathcal{C})$ is given by $k-\text{rk}(GG^\top).$  
We have two extreme cases, which will heavily impact the hardness of code equivalence: If the hull is trivial, i.e., $\mathcal{H}(\mathcal{C})= \{0\}$, then $\mathcal{C}$ is called a \emph{Linear Complementary Dual} (LCD) code. On the other hand, if $\mathcal{H}(\mathcal{C})=\mathcal{C}$, i.e.,  $\mathcal{C} \subseteq \mathcal{C}^\perp$ then $\mathcal{C}$ is called a \emph{self-orthogonal} code. 
As the next result shows, for large enough $q$ and $n$, we expect random codes to have a trivial hull.

 \begin{theorem}\cite{hull}
        Let $q$ be a prime power and $k \leq n$ be positive integers. Let $\mathcal{C}$ be a random $[n,k]_q$ linear code. Then $\mathcal{H}(\mathcal{C}) = \{0\}$ with  probability $ \geq 1 - 1/q - 1/q^2,$ for $n$ growing. 
    \end{theorem}

Several metrics can be used in coding-theoretic applications, however, in this work we will focus on the Hamming metric.
 
\begin{defi}
Let $n$ be a positive integer. For $x \in \mathbb{F}_q^n$, the support of $x$ is given by
\[
    \text{supp}(x) = \{ i \in \{1, \ldots, n\} \mid x_i \neq 0 \}.
\]
The \emph{Hamming weight} of $x$   is given by the size of its support, {i.e.},
$\text{wt}_H(x) := | \text{supp}(x) |.$
For $x,y \in \mathbb{F}_q^n$, the \emph{Hamming distance} between $x$ and $y$ is given by the number of positions in which they differ, {i.e.},
$d_H(x,y) := | \{ i \in \{1, \ldots, n\} \mid x_i \neq y_i \} |.$
The \emph{minimum Hamming distance} of $\mathcal{C}\subseteq \F_q^n$ is denoted by $d_H(\mathcal{C})$ and given by
$$d_H(\mathcal{C}) := \min \{ d_H(x, y) \mid x, y \in \mathcal{C}, \ x \neq y\}.$$
\end{defi}
If the minimum distance $d$ of an $[n,k]_q$ linear code $\mathcal{C}$ is known, we write that $\mathcal{C}$ is an $[n,k,d]_q$ linear code.
One can generalize the idea of minimum distance to the weight enumerator and weight distribution of a code:
\begin{defi}
Let $\mathcal{C} \subseteq \mathbb{F}_q^n$ be a linear code.
    For any $w \in \{0, \ldots, n\}$, let us denote by $A_w(\mathcal{C}) := |\{ c \in \mathcal{C} \mid \text{wt}_H(c) = w\}|$ the \emph{weight enumerator} of $\mathcal{C}.$
    We further denote by $(A_0(\mathcal{C}), \ldots, A_n(\mathcal{C}))$ the \emph{weight distribution} of $\mathcal{C}.$
\end{defi}

Another generalization is the so-called generalized weight. For this, we need to introduce the support of a code:

\begin{defi}
    Let $\mathcal{C}$ be an $[n,k]_q$ linear code. The \emph{support} of $\mathcal{C}$ is defined as
    $$\text{supp}_H(\mathcal{C}):= \bigcup_{c \in \calC} \text{supp} (c).$$
    The \emph{weight} of the code $\mathcal{C}$ is given by
    $\text{wt}_H(\mathcal{C}):= | \text{supp}_H(\mathcal{C})|.$\footnote{Clearly, for a non-degenerate code, the support will be full, i.e., $\{1, \ldots, n\},$ however, as soon as we go to subcodes of $\mathcal{C},$ this will change and therefore become of interest.}  Let  $r \in \{1, \ldots, k\}$.
    The $r$th \emph{generalized weight} of $\mathcal{C}$ is given by 
    $d_r(\mathcal{C}):= \min\{ \text{wt}_H(\mathcal{D}) \mid \mathcal{D} \subset \mathcal{C}, \text{dim}(\mathcal{D} )=r\} $.
\end{defi}
Note that 
 the first generalized weight $d_1(\mathcal{C})$ is the minimum distance $d_H(\mathcal{C}).$

On several occasions, it is beneficial to project the code to some coordinates, i.e., to \emph{puncture} the code in the other indices:

\begin{defi}
    Let $\mathcal{C}$ be an $[n,k,d]_q$ linear code and $S \subset \{1, \ldots, n\}$ be a subset of size $s$. The \emph{punctured code} in $S$ is
    $\mathcal{P}(\mathcal{C},S) := \{ (c_i)_{i \notin S} \mid c \in \mathcal{C}\}.$
    If we consider the subcode 
    $\mathcal{C}(S) = \{c \in \mathcal{C} \mid c_i =0 \text{ for all } i \in S\},$ then the \emph{shortened code} in $S$ is given by
    $\mathcal{S}(\mathcal{C},S):=\mathcal{P}(\mathcal{C}(S),S)= \mathcal{C}(S)_{S^C}.$
\end{defi}

    Let us denote by $*$ the componentwise product or \emph{Schur product} between $x$ and $y$, i.e.,
    $x*y= (x_1y_1, \ldots, x_ny_n)$ for $x,y \in \mathbb{F}_q^n$.  The Schur product has many algebraic properties, in particular, it is  symmetric and bilinear.  
\begin{defi}
    Let $\mathcal{C}_1$ be an $[n,k_1]_q$ linear code and $\mathcal{C}_2$ be an $[n,k_2]_q$ linear code. The \emph{Schur product code} of $\mathcal{C}_1$ and $\mathcal{C}_2$ is defined as
    $$\mathcal{C}_1 * \mathcal{C}_2 := \langle \{c_1 * c_2 \mid c_1 \in \mathcal{C}_1,c_2 \in \mathcal{C}_2\}\rangle.$$
The $\ell$th \emph{power code} of $\mathcal{C}_1$ is defined as 
    $$\mathcal{C}_1^{(\ell)} = \langle \{  c_1 * \cdots * c_\ell  \mid c_1, \ldots, c_\ell \in \mathcal{C}_1\}\rangle. $$
For $\ell=2$ we call this the \emph{square code} of $\mathcal{C}_1$.
\end{defi}

 Lastly we need the notion of (Hamming) isometries to be able to define code equivalence later on.
    A  map $\varphi: \mathbb{F}_q^n \to \mathbb{F}_q^n$ is called an \emph{isometry} if  for all $x,y \in \mathbb{F}_q^n$ we have $d_H(x,y)=d_H(\varphi(x), \varphi(y)).$

\begin{prop}[\cite{pellikaan2017codes}, Corollary 1.5.14]
The linear isometries of the Hamming metric  consist of monomial transformations, i.e., $(\mathbb{F}_q^\star)^n \rtimes   \mathcal{S}_n=:M_{n,q}$ acting on $\F_q^n$ via $((d_1, \ldots, d_n), \sigma) \cdot (x_1, \ldots, x_n) = (d_1 x_{\sigma^{-1}(1)}, \ldots, d_n x_{\sigma^{-1}(n)})$.
\end{prop}

\section{Code Equivalence}

 We will now define several types of code equivalence and the corresponding computational problems.

 \begin{defi}\label{def:equivalence}
    We say that two codes $\mathcal{C}_1,\mathcal{C}_2 \subseteq \mathbb{F}_q^n$ are \emph{linearly equivalent}, if there exists a map  $\varphi \in  (\mathbb{F}_q^\star)^n \rtimes \mathcal{S}_n$, such that $\varphi(\mathcal{C}_1)=\mathcal{C}_2$. 

  We say that two codes $\mathcal{C}_1,\mathcal{C}_2 \subseteq \mathbb{F}_q^n$ are \emph{permutation equivalent}, if there exists a permutation of indices, which transforms $\mathcal{C}_1$ into $\mathcal{C}_2$, that is, there exists $\sigma \in \mathcal{S}_n$, such that $\sigma(\mathcal{C}_1)=\mathcal{C}_2$. 
\end{defi}

Permutation equivalence is thus a special instance of the more general linear equivalence. We remark that an even more general notion is the one of \emph{semi-linear} equivalence, where also field automorphisms are taken into account. However, since the number of such field automorphisms is small (indeed $|\text{Aut}(\F_q)| = \log_p(q)$, where $p$ is the characteristic of $\F_q$), these do not play a role from a cryptographic point of view, and we will hence not consider this variant.

Naturally, we can express code equivalence in terms of matrix operations. For this let $G_1,G_2$ be the generator matrices of two $[n,k]_q$ linear codes $\mathcal{C}_1 $ and $\mathcal{C}_2$. Then there exists a $\varphi =(d, \sigma) \in (\mathbb{F}_q^\star)^n \rtimes S_n$ such that $\varphi(\mathcal{C}_1)=\mathcal{C}_2$ if and only if there exist $S \in \text{GL}_k(\mathbb{F}_q)$, $D=\text{diag}(d)$ and $P \in S_n$ such that 
    $G_2=SG_1 D P. $

Both types of equivalence behave well under duality, as the following shows.

\begin{prop}\label{prop:dualequiv}
If $\mathcal{C}_1$ and $\mathcal{C}_2$ are permutation equivalent via $\sigma \in S_n$ then $\sigma(\mathcal{C}_1^\perp)= \mathcal{C}_2^\perp.$
On the other hand, if $\mathcal{C}_1$ and $\mathcal{C}_2$ are linearly equivalent via $\varphi=((d_1, \ldots, d_n),\sigma)$ then 
$\varphi'=((d_1^{-1}, \ldots, d_n^{-1}),\sigma)$ is such that $\varphi'(\mathcal{C}_1^\perp)=\mathcal{C}_2^\perp.$
\end{prop}

A special case of equivalence arises when a code is equivalent to itself; in this case, we speak of an automorphism of the code:

\begin{defi} 
Let $\mathcal{C}$ be an $[n,k]_q$ linear code. The (monomial) \emph{automorphism group} of $\mathcal{C}$ is given by the linear isometries\footnote{Note that \emph{automorphism} sometimes also refers to any invertible linear map from $\mathbb{F}_q^n \to \mathbb{F}_q^n$ such that $\varphi(\mathcal{C})=\mathcal{C}$; however, we will assume that it is an element of $M_{n,q}.$} that map $\mathcal{C}$ to $\mathcal{C}:$
$$\text{Aut}(\mathcal{C}) := \{ \varphi \in (\mathbb{F}_q^\star)^n \rtimes S_n \mid \varphi(\mathcal{C})=\mathcal{C}\}.$$
\end{defi}

We remark that the automorphism group of a binary random linear code is with high probability trivial \cite{rigid}, i.e., $\text{Aut}(\mathcal{C})=\{\text{id}\}.$ While it is believed that for $q>2$, random codes have  automorphism group $\{\lambda \; \text{id} | \lambda \in \F_q^\star \}$, a proof is still missing in the literature.

\begin{rem}
    There are several parameters or properties of equivalent codes which remain invariant under an application of a linear isometry, e.g., the length, dimension and minimum distance. 
    Moreover, the weight enumerator, the generalized weights, and the automorphism group are invariants.\footnote{However, if two codes share the same weight enumerator or the same generalized weights, they are not necessarily equivalent, as can be seen with two Reed-Solomon codes with different evaluation points.}

From Proposition \ref{prop:dualequiv}, we also have that the hull is an invariant of permutation equivalence, i.e., if $\sigma(\mathcal{C})=\mathcal{C}'$ for some  $\sigma \in S_n$, then $\sigma(\mathcal{H}(\mathcal{C}))=\mathcal{H}(\mathcal{C}').$ 
\end{rem}

Although they are not invariants, power codes also can serve as a tool to check for equivalence, since it was shown in \cite{square} that the power codes of two linearly equivalent codes remain linearly equivalent.
\begin{lemma}
\label{lemma:power_codes_equivalent}
    If $\mathcal{C}, \mathcal{C}'$ are two $[n,k]_q$ linear codes and $\varphi=((d_1, \ldots, d_n),\sigma) \in M_{n,q}$ is such that $\varphi(\mathcal{C})=\mathcal{C}'$, then $\varphi^{(\ell)}=((d_1^\ell, \ldots, d_n^\ell), \sigma) \in M_{n,q}$ is such that 
    $\varphi^{(\ell)}(\mathcal{C}^{(\ell)} )= \mathcal{C}'^{(\ell)}.$
\end{lemma}

\subsection*{Code Equivalence Problems}

We now define the various code equivalence problems that appear in cryptographic contexts. For this note that for complexity classifications the decisional versions are used, while for actual cryptosystems we usually turn to the computational versions of these problems. We will now formulate the computational ones, where we start with the Linear and the Permutation Equivalence Problems (LEP and PEP), according to the two types of equivalence from Definition \ref{def:equivalence}.

\begin{problem}[LEP]
    Given two $[n,k]_q$ linear codes $\mathcal{C}, \mathcal{C}'$ find (if it exists) a $\varphi \in  (\mathbb{F}_q^\star)^n \rtimes S_n$    such that 
    $\varphi(\mathcal{C})=\mathcal{C}'.$
\end{problem}

\begin{problem}[PEP]
    Given two $[n,k]_q$ linear codes $\mathcal{C}, \mathcal{C}'$ find (if it exists) a $\sigma \in  S_n$    such that 
    $\sigma(\mathcal{C})=\mathcal{C}'.$
\end{problem}

As before, PEP is a special instance of LEP, with the scalars all being equal to one. In a similar spirit one could also allow the scalars to be arbitrary, but fix the permutation to be the identity, which would result in the following question:  given two generator matrices $G,G' \in \mathbb{F}_q^{k \times n}$ find $d \in (\mathbb{F}_q^\star)^n$ and $S\in \mathrm{GL}_k(\F_q)$ such that $SG \text{diag}(d) =G'$. However, this problem can easily be solved by computing the reduced row echelon forms of both generator matrices and then solving a system of $k(n-k)$ linear equations in $n-k$ variables. Therefore, we do not consider this variant in cryptographic applications.

While PEP and LEP are the main problems considered in the literature, there are several variants, which are captured by the following problem \cite{battagliola2026powerpowercodesnew}:

\begin{problem}[LEP($U$)]
Let $U \subseteq \mathbb{F}_q^\star$ be a multiplicative subgroup. Given two $[n,k]_q$ linear codes $\mathcal{C}$ and $\mathcal{C}'$   find (if it exists) an isometry $\varphi=(d, \sigma) \in U^n \rtimes S_n$ such that $\varphi(\mathcal C)=\mathcal C'$.
\end{problem}
By setting $U= \mathbb{F}_q^\star$, we recover LEP, for $U=\{1\}$, we get PEP and for $U= \{ \pm 1\}$, we get the  Signed Permutation Equivalence Problem (SPEP).
Note that in the Lee metric, the linear isometries are exactly the signed permutations. 
For $q=2$, LEP is PEP, and for $q=3$, LEP is SPEP.

Another generalization of PEP considers subcodes. It is also known as \emph{Permuted Kernel Problem} (PKP) (see e.g. \cite{paolopkp}), and is used in the signature scheme PERK \cite{perk}:
 
 \begin{problem}[PKP]
Let $0<k' \leq k \leq n$. Given an $[n,k]_q$ code $\mathcal{C}$ and an $[n,k']_q$ code $\mathcal{C}'$, find (if it exists) a $\sigma \in S_n$, such that $\mathcal{C}' \subseteq \sigma(\mathcal{C}).$

 \end{problem}

While it was shown in \cite{rothGI} that the decisional version of PKP is NP-hard, the (decisional) variants of the code equivalence problems generalized by LEP($U$) lie in the complexity class $\mathrm{AM} \cap \mathrm{coAM}$ \cite{rothGI}. As for other isomorphism-type problems (e.g.\ graph isomorphism), this suggests that they are unlikely to be NP-hard, since NP-hardness would imply a collapse of the polynomial hierarchy. While this is shown in \cite{rothGI} for LEP, the argument extends directly to LEP$(U)$ for any multiplicative subgroup $U \subseteq \mathbb{F}_q^\star$.
Despite this, no efficient algorithm is known for solving code equivalence in general.

 \section{Reductions to Other Problems}

One way of solving any of the code equivalence problems is to reduce it to another problem for which a (efficient) solving algorithm is known. In this section we give an overview of various such reductions.

\subsection{Linear Equivalence to Permutation Equivalence}

There exists a very elegant reduction from linear equivalence to permutation equivalence \cite{hull}. For this the authors introduce the \emph{closure} of a code.
We describe a generalization from \cite{battagliola2026powerpowercodesnew}.

\begin{defi}
    Let $U \subseteq \mathbb{F}_q^\star$ be a multiplicative subgroup.
   Let $\mathcal{C}$ be an $[n,k]_q$ linear code and $\alpha \in \mathbb{F}_q$ be a primitive element.
   Let $r = |U|$ such that $r \mid (q-1)$.
   Then $U$ is the subgroup generated by $\alpha^{(q-1)/r}$.
Define    $\lambda  = (1, \alpha^{(q-1)/r}, \dots, \alpha^{(r-1)(q-1)/r})$.
The $r$-th \emph{partial closure} of $\calC$ is given by the Kronecker product $ \lambda \otimes \mathcal{C}$.
In the case $U = \F_q^\star$ we just speak of the \emph{closure}.
\end{defi}

The new code is now of length $n r$ and still of dimension $k.$ In fact, if $G$ is a generator matrix of $\mathcal{C},$ then $\lambda \otimes G$ is a generator matrix of $\lambda \otimes \mathcal{C}$ \cite{hull}.

\begin{prop}[\cite{battagliola2026powerpowercodesnew}]
    Let $\mathcal{C}_1, \mathcal{C}_2$ be two $[n,k]_q$ linear codes. Then $\calC_1$ is equivalent in the $\text{LEP}(U)$-sense to $\calC_2$ if and only if $\lambda \otimes \mathcal{C}_1$ is permutation equivalent to $\lambda \otimes \mathcal{C}_2$.
    In particular we get a reduction from $\text{LEP}(U)$ to PEP in time $\poly(n, |U|)$.
\end{prop}
This effectively transforms LEP($U$) into a PEP instance of length $rn$.
The first case of such a reduction in case $U = \F_q^\star$ was done in \cite{hull}.
Also the special case $U = \{\pm 1\}$, i.e., the reduction from SPEP to PEP has already been done before in \cite{ducas2023hull}.
However, the next proposition which is a generalization of \cite{hull} shows that the partial closure is usually self-orthogonal.
\begin{prop}
\label{prop:partial_closure_self_orthogonal}
      If $|U| > 2$ then the partial closure $\lambda \otimes \mathcal{C}$ is self-orthogonal.
      In particular, for $q \geq 4$ the closure is self-orthogonal.
\end{prop}
       
This result relies on the fact that $\sum_{\beta \in U} \beta^2=0$ if squaring defines a non-trivial character.

\subsection{Permutation Equivalence and Graph Isomorphism}\label{sec:adj}

For this reduction, let us first recall some graph theory. 
A graph $\mathcal{G}$ consists of vertices $V$ and edges $E$ between the vertices, i.e., $E\subset V\times V.$
We will focus on undirected graphs, thus whenever $(u,v) \in E$ also $(v,u) \in E$ and we label the edges with a weight $w(u,v)$.
We say that two weighted graphs $\mathcal{G}=(V,E)$ and $\mathcal{G}'=(V',E')$ are isomorphic, if there exists a bijective map $f:V\to V'$ with $ (u,v) \in E$ if and only if $(f(u), f(v)) \in E'$ and  $w(u,v)= w(f(u),f(v))$.
Thus, we can focus on $V=V'=\{1,\ldots,n\}$ and maps $f=\sigma \in \mathcal{S}_n$.

\begin{problem}[Weighted Graph Isomorphism Problem]
Given $\mathcal{G},  \mathcal{G}'$, find (if it exists) isomorphism $\sigma \in \mathcal{S}_n$ mapping $\mathcal{G}$ to $\mathcal{G}'.$
\end{problem}

There are several reductions from graph isomorphism to permutation equivalence or linear equivalence including \cite{rothGI, kaski2006classification, Grochow12, bennett2025improvements}.
However, from here on we are interested in the converse direction.

The \emph{adjacency matrix} of a weighted graph $\mathcal{G}$ is defined as the $n \times n$ matrix $A$ with entries  $A_{i,j}=w(i,j)$ if $(i,j) \in E$ and $0$ else.
Note that two graphs $\mathcal{G}, \mathcal{G}'$ are isomorphic if and only if there exists a permutation matrix $P$ such that $P^\top A P= A'.$

The GI problem has been shown to be quasi-polynomial time by Babai in 2017 \cite{babai}, that is we can solve GI in time $2^{O(\log(n)^c)}$ for some constant $c$.
This makes the following randomized reduction \cite{magali} very interesting:
For $\mathcal{C}$ with trivial hull, we define 
$$A(G) = G^\top (G G^\top)^{-1} G.$$
Clearly, this matrix can only exist if $GG^\top$ is invertible, i.e., if the hull of $\mathcal{C}$ is trivial. The matrix $A(G) \in \mathbb{F}_q^{n \times n}$ is symmetric, $\langle A(G)\rangle = \mathcal{C}$ and, moreover, it is independent of the choice of $G,$ as $A(SG)=A(G)$ for $S\in\text{GL}_k(\mathbb{F}_q).$

\begin{theorem}\cite{magali}
    PEP can be reduced to weighted GI, for codes with trivial hull.
\end{theorem}

This reduction of PEP to GI can easily be extended to SPEP. In fact, if $G'=SGDP$, for some $S \in \text{GL}_k(\mathbb{F}_q),$ permutation matrix $P$ and $D=\text{diag}(d)$ and $d \in \{\pm 1\}^n$, then 
$A(G')= P^\top D A(G) DP, $
which is another version of GI, called signed graph isomorphism, which is  GI-complete and hence the  solver of Babai can also be used to solve SPEP, whenever $GG^\top$ is invertible.

For LEP it would be natural to apply the reduction via the closure to PEP and then to GI. 
However, this does not  work for $q>3$, due to Proposition \ref{prop:partial_closure_self_orthogonal}.

     The case $q=4$ can be handled using the \emph{Hermitian inner product} instead of the
standard one \cite{hull}. This product exists in general when $q=p^{2m}$ and is defined as $
\langle x,y\rangle_H :=\sum_{i=1}^n x_i y_i^{p^m}$ for any $x,y\in\mathbb{F}_q^n$. 
This form is $\F_{p^m}$-linear in each component and satisfies
$\langle x,y\rangle_H =(\langle y,x\rangle_H)^{p^m}$.

Hence, we define the Hermitian dual by
$
\mathcal C^\star :=\{x\in\mathbb F_q^n \mid \langle x,y\rangle_H=0
\text{ for all } y\in\mathcal C\}.
$
Let $G^{[p^m]}, H^{[p^m]}$ be the matrices given by raising each element to the $p^m$-th power.
If $\mathcal C=\langle G\rangle=\ker(H^\top)$, then
$
\mathcal C^\star=\langle H^{[p^m]}\rangle=\ker((G^{[p^m]})^\top)$ and $ (\mathcal C^\star)^\star=\mathcal C.$
The \emph{Hermitian hull} is defined as $\mathcal H^\star(\mathcal C):=\mathcal C\cap\mathcal C^\star.$
If the Hermitian hull is trivial, we can define $$
A^\star=(G^{[p^m]})^\top\!\left(G(G^{[p^m]})^\top\right)^{-1}G,$$
which is again independent of the generator matrix.
As before, random codes have trivial Hermitian hull with high probability
\cite{hull}. It remains to check, whether closures also have trivial hulls: Since $$ (\lambda\otimes G)((\lambda\otimes G)^{[p^m]})^\top=\lambda(\lambda^{[p^m]})^\top\,G(G^{[p^m]})^\top,$$ we are left with computing
$ \lambda(\lambda^{p^m})^\top
=\sum_{\beta\in\mathbb F_q^\ast}\beta^{p^m+1}, $
which is only non-zero if $(p^{2m}-1)\mid(p^m+1)$,  which implies
$p=2,m=1$, i.e., $q=4$. This shows that this approach cannot be generalized to larger $q$.

\subsection{Search to Decision}

While it is easy to see that the decisional version of any code equivalence problem can be reduced to the search (or computational) version, the other direction is not obvious. 
In \cite{giacomo}, the authors prove a search-to-decision reduction for
PEP:\footnote{For LEP this is still an open question.}  

Suppose we are given an oracle
that decides whether two linear codes are permutation equivalent. The
reduction recovers the secret permutation by iteratively determining the
image of each coordinate. 
More precisely, let $\mathcal{C}, \mathcal{C}'$ be two codes and $\sigma \in S_n$ be such that  $\sigma(\mathcal{C})=\mathcal{C}'$. For an index $i$  the reduction
tests whether $\sigma(i)=j$ by appending multiple copies of selected columns to the
generator matrices of $\mathcal{C}$ and $\mathcal{C}'$. These repetitions enforce that any permutation
on the augmented codes must have $\sigma(i)=j$.
Querying the oracle on the augmented codes therefore decides
whether $\sigma(i)=j$. By repeating this procedure a polynomial number of times, one recovers $\sigma$.

A big open question is whether such an (efficient) oracle exists, e.g., an invariant which is characterizing and efficient to compute.

\subsection{SPEP to LIP and additional reductions}

\begin{problem}[Lattice Isomorphism Problem (LIP)]
    Given two lattices $\Lambda_1, \Lambda_2$, find (if it exists) an orthogonal matrix $O$ such that
 $       \Lambda_1 = \Lambda_2 O.
    $ \end{problem}
It is well known that this can be rephrased in terms of quadratic forms.
In \cite{bennett2025relating} the authors prove a reduction from Signed Permutation Equivalence for prime fields with cardinality $p$ ($\SPEP_p$) to Lattice Isomorphism  Problem (LIP).
\begin{theorem}[\cite{bennett2025relating}]
    Let $p$ be a prime and $n \in \Z_{>0}$.
    There is a $\poly(n, \log(p))$ reduction from $\SPEP_p$ on codes with  length $n$ and minimum distance at least $p^2 + 1$ to LIP on lattices of rank $n$.
\end{theorem}
The basic idea is as follows:
From a code $\calC$ we can use the Construction A lattice $\calC + p \Z^n$ to get a lattice.
One can show that the minimum distance requirement ensures that the shortest vectors in this lattice are of the form $\pm p e_i$ where $e_i$ is a standard basis vector.
Since orthogonal matrices preserve lengths and are invertible, in this case the isomorphisms between two such Construction A lattices have to be signed permutations.

Additionally, the authors present a reduction from LEP to LIP via several intermediate steps, one of which involves UCE, which we do not introduce here for brevity. In particular, the use of closure in one of these steps renders the reduction less efficient, yielding only a $\poly(n,p)$ complexity. The authors also provide improved reductions for some previously known results. We summarize the relevant reductions below.

\begin{itemize}
    \item  
    There is a reduction from PEP to LEP that runs in $\poly(n, \log(q))$ time \cite{cheraghchi2025reductions}. (With the results from above this implies a reduction from PEP to LIP.)
    \item There is a reduction from PEP to SPEP  in $\poly(n, \log (q))$ time \cite{cheraghchi2025reductions}.
    \item  There is a reduction from LEP to Matrix Code Equivalence (MCE) that runs in $\poly(n)$ time \cite{rank}.
    For one of the two reductions from LEP to MCE of \cite{rank} the authors simply embed a vector in a diagonal matrix.
    \item 
    Using ideas of  \cite{Mathon79} and combining the results of \cite{grochowautom20} and \cite[Theorem 5]{bardetbrionotmanisaeedsendriercodewordmatching26} one  gets a probabilistic reduction from PEP and LEP to the problem of finding the order of the (permutation or monomial) automorphism group of a code.
\end{itemize}

\begin{figure}
    \centering
    \begin{tikzpicture}[
    node distance=1.5cm and 1.5cm,
    every node/.style={font=\large},
    >={Stealth[length=5pt]}
]

\node (GI)   {GI};
\node (LEP)  [right=of GI]   {LEP};
MCE\node (MCE) [above right=1.6 and 1.45 of GI]  {MCE}; 
\node (UCE)  [right=of LEP]  {UCE};
\node (LEPU) [above right=1.5 and 1.2cm of UCE] {LEP(U)};
\node (PEP)  [above right=0.3cm and 1.5cm of UCE]  {PEP};
\node (SPEP) [below right=0.7cm and 1.43cm of UCE]  {SPEP};
\node (LIP)  [above right=0.0cm and 1.5cm of {$(PEP)!0.5!(SPEP)$}] {LIP};

\draw[->] (GI) -- (LEP);

\draw[->] (LEP) -- (MCE);

\draw[<->] (LEP) -- (UCE);

\draw[<->] (LEP) -- (PEP);

\draw[->] (UCE) -- (PEP);

\draw[->] (UCE) -- (SPEP);

\draw[<->] (PEP) -- (SPEP);

\draw[->] (SPEP) -- (LIP);

\draw[->] (LEPU) -- (PEP);

\end{tikzpicture}
    \caption{Known reductions between several equivalence and isomorphism problems.}
    \label{fig:reductions}
\end{figure}
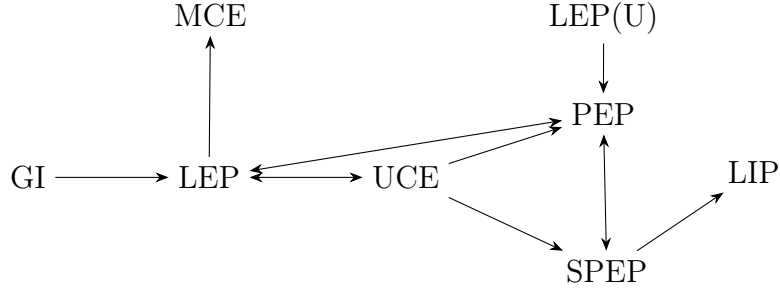

We summarize the known reductions in Figure~\ref{fig:reductions}, where we omit certain reductions that can be derived through alternative approaches.

A natural direction for future work is to extend existing reductions to encompass LEP($U$). While some of these extensions may be relatively direct, a systematic treatment remains open.
We emphasize that the figure does not reflect the parameter dependencies of the reductions. In particular, some run in time polynomial in $n$ and $q$, while others are polynomial in $n$ and $\log(q)$. Achieving reductions that uniformly run in time $\poly(n, \log(q))$ would be desirable.
Moreover, the reductions to LIP are currently restricted to prime fields, indicating a further limitation.
Finally, additional reductions have been established in the literature~\cite{grochow2023complexity, d2024monomial, kreuzer2025code}, but fall beyond the scope of this survey.

 \section{Solvers}\label{sec:solver}
In this section we will give an overview of the known solvers for the different code equivalence problems. The computational complexities will be  summarized in Tables \ref{tab:summarytable1} and \ref{tab:summarytable2} in Section \ref{sec:summary}.

 \subsection{Algebraic Solvers}

The algebraic solver goes back to Saeed's thesis \cite{saeed}, where he uses that if $ \mathcal{C}= \langle G \rangle = \text{ker}(H^\top)$ and $\mathcal{C}'= \langle G'\rangle = \text{ker}(H'^\top)$ are linearly equivalent, i.e., there exist $S \in \text{GL}_k(\mathbb{F}_q), D= \text{diag}(d), P \in S_n$ such that $SGDP= G'$,  then
$G'H'^\top =0$  implies that $ GDPH'^\top =0.$
Hence it is enough to solve the linear system $G M H'^\top =0,$ for the monomial matrix $M=DP$, with $k(n-k)$ equations and  where the entries of $M$ are the $n^2$ unknowns. The main problem is: how to put the fact that we are only interested in $M=DP$ into this model as equations?

\subsection{Combinatorial Solvers using Invariants}
The combinatorial solvers date back to Leon \cite{leon}, and have been improved by Beullens \cite{beullens2020not}.
This line of combinatorial solvers is often referred to as \emph{codeword search}, to separate it from the canonical form solvers in the next section.
In \cite{grochowautom20} it was shown that for some classes of codes, knowing the automorphism group of $\mathcal{C}$ is enough to solve LEP. This also underpins the idea of Leon \cite{leon}.

The idea of the codeword-search solvers is to find  subsets of codewords $S \subset \mathcal{C}$ and $S' \subset \mathcal{C}'$ which are also invariant under the secret monomial $\varphi,$ i.e., $\varphi(S)=S'.$
For the smaller sets $S,S'$ it will become easier to find an isometry between them.  This is usually done through invariants. However, to find the subset $S$  most solvers first need to  solve a much harder problem: the problem of finding low weight codewords, which is known to be NP-hard. Thus, their bottleneck is usually the exponential cost of an Information Set Decoding (ISD) algorithm.   

 \textbf{Leon:} The main observation of Leon \cite{leon} is that the sought isometry has to map all codewords of  weight  $w$ in $\mathcal{C}$ to all codewords of   weight $w$ in $\mathcal{C}'$. Thus, Leon first constructs the sets 
\begin{align*} 
S & = \{ c \in \mathcal{C}\mid \text{wt}_H(c)=w \}, \quad  S'  =  \{ c' \in \mathcal{C}'\mid \text{wt}_H(c')=w \}  \end{align*} and then searches for $\varphi \in M_{n,q}$ such that $\varphi(S)=S'.$
 The cost of finding such a map $\varphi$ is polynomial in the size of $S$. 
However, to construct the sets, we rely on ISD.

\textbf{Beullens' Algorithm for PEP:}  The main observation of Beullens \cite{beullens2020not} is that  a permutation $\sigma$ does not only fix the weight of a vector, but also the multiset of its entries. 
Thus, instead of seeing the elements in $S$ and $S'$ as vectors $c$ respectively $c'$, Beullens' algorithm treats them as multisets   and searches for a collision in $S \times S'$, i.e., $c \in S$ has the same multiset as $c'\in S'$.
This allows us to store less elements in $S$.
Each found collision is then used to piece-wise reconstruct the  permutation: if $c, c'$ have the same  multisets and $c_i \neq c'_j$,
then we guess $\sigma(i) \neq j$.
 Again, the bottleneck remains constructing the sets $S, S'$ using ISD.

\textbf{Beullens' Algorithm for LEP:} Beullens also provides an algorithm to solve LEP \cite{beullens2020not}. In the case of linear equivalence, the multiset of the entries are clearly not preserved. However, for any subcode $\mathcal{D} < \mathcal{C}$ there exists a $\varphi(\mathcal{D})=\mathcal{D}'< \mathcal{C}'$ with the same dimension and support size.  
In particular, Beullens proposes to use subcodes of dimension 2, that is: Beullens uses as invariant the second generalized weight.  The lists now contain generator matrices in $\mathbb{F}_q^{2 \times n}$: \begin{align*} S& =\{ D \in \mathbb{F}_q^{2 \times n} \mid \langle D \rangle < \mathcal{C}, |\text{supp}(\langle D \rangle)|= s\},  \end{align*} and similarly we construct $S'.$
One then again searches for collisions, i.e., an isometry $\varphi(\langle D\rangle)= \langle D' \rangle.$

There is also another combinatorial solver introduced by Sendrier \cite{SSA}, the so-called Support Splitting Algorithm (SSA).

\textbf{SSA:} This algorithm solves PEP in a runtime of $\mathcal{O}(q^{\text{dim}(\mathcal{H}(\mathcal{C}))})$,  
under heuristic assumptions on the refinement process, and thus not reflecting worst-case complexity.
As random codes have with high probability $\text{dim}(\mathcal{H}(\mathcal{C}))=c$ for some small constant $c$, this algorithm already breaks many instances and suggests that the hardest instances of PEP are self-orthogonal codes.

The idea of SSA is to first puncture the codes $\mathcal{C}$ and $\mathcal{C}'$ in each index $i \in \{1, \ldots, n\},$ getting the lists 
$$\mathcal{C}_1=\mathcal{P}(\mathcal{C}, \{1\}), \ldots, \mathcal{C}_n=\mathcal{P}(\mathcal{C}, \{n\}), \quad \mathcal{C}_1'=\mathcal{P}(\mathcal{C}', \{1\}), \ldots, \mathcal{C}_n'=\mathcal{P}(\mathcal{C}', \{n\}).$$

The algorithm then heavily relies on the fact that the hull is an invariant for permutation equivalence, which is also the main hindrance in generalizing this algorithm for LEP. 

In fact, it then computes the hulls $\mathcal{H}(\mathcal{C}_i)$ and $\mathcal{H}(\mathcal{C}_j')$ for all $i,j \in \{1, \ldots, n\}$. As the hulls are small, we can even compute their weight distributions and (hopefully) they reveal whether there exists a unique $\sigma \in S_n$ such that 
$\sigma(\mathcal{H}(\mathcal{C}_i))= \mathcal{H}(\mathcal{C}_j')$ for all $i,j \in \{1, \ldots, n\}$ which can also be extended to a $\pi \in S_n$ such that $\pi(\mathcal{C})=\mathcal{C}'.$ If there are, however, different choices, i.e., there exists several permutations which would send $i$ either to $j$ or to another $j'$, we continue the puncturing process, until we get a unique permutation.

In \cite{SSA}, Sendrier also mentions the possibility of extending or shortening the code. For this we recall that shortening the dual is the same as the dual of the punctured code, i.e., $\mathcal{S}(\mathcal{C}^\perp,J) = (\mathcal{P}(\mathcal{C},J))^\perp$. Hence 
$\mathcal{H}(\mathcal{P}(\mathcal{C},J))= \mathcal{H}( \mathcal{S}(\mathcal{C}^\perp, J))$.

\subsection{New Solvers using Power Codes}

While the square code is primarily used in cryptanalysis and for distinguishing structured systems, it also turns out to be a valuable tool for code equivalence.

In fact, in \cite{square}, the authors showed that we can also employ the square code to solve  (signed) permutation equivalence and in turn attacked the proposal \cite{albrecht2025hollow}.

  We can use the square of the hull to solve PEP faster: for this consider two  permutation equivalent codes $\mathcal{C}, \mathcal{C}'$, which are self-orthogonal. 
The idea of \cite{square} is to first puncture $\mathcal{C}$ and $\mathcal{C}'$ in the positions $I$, respectively $I'$, denoted by $\mathcal{P}(\mathcal{C},I)$ and $\mathcal{P}(\mathcal{C}',I')$, and then to compute the square codes of their hulls. These are again permutation equivalent. 

Thus, a first approach  for solving linear equivalence would be to raise $\mathcal{C}$ and $\mathcal{C}'$ to the power $(q-1)/2$. 
The main problem of this idea  is that by computing the $(q-1)/2$-th  power code, we get  that the dimension quickly reaches the length $n$, that is 
$  \binom{k+(q-1)/2-1}{(q-1)/2}\geq n.$

However, in the case that $n>\binom{k+(q-1)/2-1}{(q-1)/2},$ the $((q-1)/2)$-th power code approach can be applied and, starting with linearly equivalent codes, Lemma \ref{lemma:power_codes_equivalent} yields power codes that are equivalent with a signed permutation.
If the hull of these power codes is small enough, we can use SSA to solve this.
Assuming that we found the correct permutation, we can then solve LEP for the original code.

This idea can be improved, as done in \cite{battagliola2026powerpowercodesnew}.
For this, the authors generalize
the adjacency-matrix construction from Section \ref{sec:adj} to pairs of codes.
Let $A, B$ be generator matrices of linearly equivalent codes $\mathcal{A}, \mathcal{B}$.
The authors define auxiliary codes $\mathcal{A}_1, \mathcal{A}_2, \mathcal{B}_1, \mathcal{B}_2$ with generator matrices $A_1, A_2$ coming from $A$ and $B_1, B_2$ coming from $B$.
Then they define
\(
\text{Adj}(A_1, A_2)
:= A_2^\top (A_1 A_2^\top)^{-1} A_1,
\)
and similarly $\text{Adj}(B_1, B_2)$.
The auxiliary codes should force the diagonal scalings to cancel.
The most basic definition of the auxiliary codes is to use $\mathcal{A}_1 = \mathcal{A}_2 = \mathcal{A}^{(\frac{q-1}{2})}, \mathcal{B}_1 = \mathcal{B}_2 = \mathcal{B}^{(\frac{q-1}{2})}$, as described above.
However, we may also use Frobenius powers for non-prime fields.
For these auxiliary codes, they compute $\text{Adj}(A_1, A_2), \text{Adj}(B_1, B_2)$, concluding equivalence if the multisets of their diagonal entries are identical.
Note that even though it is not guaranteed that this test avoids false positives - it does avoid false negatives - the probability that this happens becomes negligible when the dimension of the auximilary codes $\mathcal{A}_i, \mathcal{B}_i$ satisfies $k^\prime \ll n.$

This yields several new classes of weak LEP instances, in particular for
small-rate codes, and provides the first attack on LEP exploiting explicitly
algebraic properties of the underlying field.

\subsection{Canonical Forms}

A recent line of work studies LEP using
\emph{canonical forms}. Instead of directly searching for the secret
monomial transformation, the idea is to
associate to each code a canonical representative of its equivalence
class. If two codes are equivalent, then their canonical representatives
must coincide. This approach was introduced  in \cite{chou2025linear}, where the authors also managed to use this new result to get smaller sizes for LESS, and was later improved by \cite{nowakowski2025improved}. 

These solvers all stem from the following observation: if $I$ is an information set for $\mathcal{C}$, and we know the secret permutation in $I$, i.e., $\sigma_I$, we can find the whole monomial transformation in polynomial time, see e.g. \cite{budroni2025two}.  Even more, recall that the hardest instances of PEP are self-orthogonal codes. 
Due to \cite{square}, we know that puncturing a self-orthogonal code in an information set results in an LCD code.
Thus, given two linearly equivalent codes, one may first compute their closures, then puncture in the corresponding information sets $I$ and $\sigma(I)$, and finally apply the reduction from PEP with trivial hull to GI.

In \cite{chou2025linear}, they take this fact even further and introduce the notion of
\emph{Left-Right Linear (LRL) equivalence}: two matrices
$ G_1=\begin{pmatrix} \text{Id}_k & A_1 \end{pmatrix},  G_2= \begin{pmatrix} \text{Id}_k & A_2 \end{pmatrix}$
are LRL-equivalent if there exist monomials
$Q_r \in M_{k,q}$ and $Q_c \in M_{n-k,q}$ such that
$A_2 = Q_r A_1 Q_c.$
Thus the  matrices $A_1,A_2$ are equivalent up to monomial row and column
transformations.

A \emph{canonical form function} takes $G= \begin{pmatrix} \text{Id}_k & A \end{pmatrix}$ as input and
outputs a canonical representative $G^*=\begin{pmatrix} \text{Id}_k & A^* \end{pmatrix}$ of its
LRL-equivalence class together with monomials satisfying
$A^*=Q_r A Q_c$, or an error symbol $\perp$.
If the algorithm succeeds on two LRL-equivalent inputs, it outputs the
same representative.  The original canonical form construction achieves constant success
probability only for large fields, whereas for constant $q$, the
success probability becomes exponentially small in $n$.
This limitation was addressed in \cite{nowakowski2025improved}, where the author introduced an improved
canonical form function with success probability $1-O(n^{-1})$ for all
$q \ge 7$.

\subsubsection{Further Improvements}
 
In \cite{bennett2025improvements}, the authors present asymptotic improvements for several algorithms related to code equivalence. They give a faster deterministic algorithm for LEP, extending Babai’s approach from \cite{bcgq11} beyond the permutation case, and avoiding the exponential length expansion of the classical closure by handling monomial isometries directly. They also propose a randomized algorithm for LEP and PEP based on a meet-in-the-middle strategy combined with a down-up random walk for near-uniform matroid basis sampling. Finally, they introduce a quantum variant using the BHT collision-finding framework based on Grover's algorithm. The latter two algorithms achieve runtimes comparable to \cite{nowakowski2025improved}, while applying to arbitrary codes and not requiring $q \geq 7.$

\subsection{Solvers using Side Information}
Another line of work studies LEP under side information, namely access to pairs of equivalent codewords $(c,c') \in \mathcal{C} \times \mathcal{C}'$ connected through the secret monomial.
Such pairs naturally arise when computing lists of low-weight codewords, and earlier approaches required $\Omega(\log n)$ pairs to recover the monomial transformation, while \cite{budroni2025two} showed that two pairs suffice with high probability by exploiting support intersections and ratio structures. 
More recently, \cite{onepair} and \cite{bardetbrionotmanisaeedsendriercodewordmatching26} demonstrated that even a single pair can be sufficient, achieving the theoretical lower bound of this framework. 
A complementary direction considers the reuse of the same secret monomial across multiple instances, where several equivalent pairs induce linear constraints on the underlying permutation. 
As shown in \cite{twice}, these constraints can be exploited to distinguish shared secrets and partially recover the permutation, highlighting the risks of monomial reuse in cryptosystems.

An open question which remains, is which kind of information can be obtained through side-channel attacks and how will they influence the cost of LEP solvers?

\subsection{Summary of the Costs}\label{sec:summary}
Tables \ref{tab:summarytable1} and \ref{tab:summarytable2} provide a short summary on the costs of the previously presented solvers. We use the following notation: $H$ for the binary entropy function.
$N_w$ denotes the number of codewords of weight $w$  and $C_{\text{ISD}}$ denotes the exponential cost of an ISD algorithm of choice.

 \begin{table}[ht]
\begin{tabular}{p{4cm}p{4.1cm}p{3cm}}
\textbf{Algorithm} & 
\textbf{Complexity} & \textbf{Remarks} \\\hline

Brute Force
& $2^{k \log n +o(k \log_2 n)}$ 
& Guess  $\sigma$ on an information set \\ \hline 

Via Graph Isomorphism
& $2^{n + o(n+q)}$\cite{bennett2025improvements}
&  Run Babai's algorithm for each information set, i.e., $\binom{n}{k}$ times\\\hline

Meet-in-the-middle variant
& $2^{n/2+o(n+q)}$\cite{bennett2025improvements}
& Extension of GI-approach, probabilistic, randomized \\\hline

Algebraic (Gröbner basis)
& $2^{n \log \frac{k(n-k)}{n^2} + o(n \log \frac{k(n-k)}{n^2})}$\cite{saeed}
& Heuristic
\\\hline

SSA
& $\mathcal O (n^3 +q^hn^2 \log n)$\cite{magali,hull}
& Heuristic,  $h$ is hull dimension  \\\hline
\end{tabular}
\caption{Complexities of various solvers for PEP.}
\label{tab:summarytable1}
\end{table}

 \begin{table}[ht]
\begin{tabular}{p{4cm}p{4.1cm}p{3cm}}
\textbf{Algorithm} & 
\textbf{Complexity} & \textbf{Remarks} \\\hline

Brute Force & $2^{k \log n +o(k \log_2 n)}$  
& Guess $\sigma$ on an information set (as diagonal can be solved once $P$ is known) \\ \hline 

Via Graph Isomorphism 
& $2^{2(q-1)n + o(n)}$\cite{bennett2025improvements}
&  Run Babai's algorithm $\binom{n}{k}$ times on the code closure\\\hline

Meet-in-the-middle variant
& $2^{n/2+o(n+q)}$\cite{bennett2025improvements}
& Extension of GI-approach, probabilistic\\\hline

Leon’s Algorithm 
& $\mathcal O(\log(N_w) C_{ISD})$\cite{paolo}
& ISD based\\\hline

Beullens' Algorithm  
& $\mathcal O(\sqrt{\frac{n \log n}{N_w}} C_{ISD})$\cite{paolo}
& ISD based   \\\hline

Nowakowski's Algorithm 
& $2^{n/2 H(k/n) + o(n)}$\cite{nowakowski2025improved}
& Randomized, $q\geq 7$\\\hline

SSA extension 
& $\mathcal{O} (n^3 +q^k n^2 \log n)$ \cite{hull}
& SSA applied to closure of code \\ \hline

\end{tabular}
\caption{Complexities of various solvers for LEP.}
\label{tab:summarytable2}
\end{table}

\section{New Directions and Open Questions}\label{sec:conclusion}

\paragraph{Every Code is Equivalent to a Code with Trivial Hull:}
An interesting discovery has been made in \cite{carlet}: every linear code over $\mathbb{F}_q$ with $q>3$ is linearly equivalent to a code with trivial hull, i.e., an LCD code.
Unfortunately, this cannot be combined with the existing reductions: if we first reduce LEP to PEP by using the closure, we  get self-orthogonal codes, which are permutation equivalent. If we now apply   \cite{carlet}, we end up with LCD codes, but they are again linearly equivalent and not permutation equivalent.

\paragraph{The Search for New Codes Respecting the Monomial Equivalence:}
One of the open questions is whether there exists another code $F(\mathcal{C})$, which acts for LEP like the dual code for PEP. That is: for a generator matrix of a linearly equivalent code, $G'=SGDP$ with $S \in \text{GL}_k(\mathbb{F}_q),$ a permutation matrix $P$ and diagonal matrix $D$, we want that $F(H')= S' F(H) D^{-1}P$, for some $S' \in \text{GL}_k(\mathbb{F}_q)$ and, to use the reduction to GI, we additionally require that $G F(H)^\top$ is invertible.
In fact, if such $F(\mathcal{C})$ construction exists, we have that
$\text{Adj}(G',F(H'))=  P^\top D \text{Adj}(G,F(H)) DP,$ yet another version of GI, which is still GI-complete and hence, we can apply the quasi-polynomial time solver of Babai to recover $(P,D).$ One such example was the Hermitian hull. Other inner products have, unfortunately, not led to such a reduction.

\paragraph{Rank-Metric Code Equivalence is Easy:}
In \cite{rank} the authors state that the code equivalence problem for $\mathbb{F}_{q^m}$-linear rank-metric codes is either in $P$ or in $ZPP$, depending on the size of the base field.
More precisely, the $\F_{q^m}$-linear rank-metric code equivalence problem is: given two generator matrices $G,G' \in \mathbb{F}_{q^m}^{k \times n}$, find $S \in \text{GL}_k(\mathbb{F}_{q^m})$ and $P \in \text{GL}_n(\mathbb{F}_q)$ such that $SGP=G'.$ As this problem is easier than its Hamming-metric counterpart, a natural question is whether we may get a reduction from LEP to the $\mathbb{F}_{q^m}$-linear version of rank-metric equivalence, e.g. along the lines of \cite{random}.

\paragraph{Extension Fields: Some New Approaches:}

If LEP is defined over $\mathbb{F}_{p^s}$ with $s>1$, a natural idea is to descend to
the base field $\mathbb{F}_p$ and try to solve the resulting instance there. Two
standard ways to do so are via the \emph{expanded code} and the \emph{trace
code}.
Fix a basis $\Gamma=\{\gamma_0,\ldots,\gamma_{s-1}\}$ of $\mathbb{F}_{p^s}$ over
$\mathbb{F}_p$. Writing each $x\in \mathbb{F}_{p^s}$ as $x=\sum_{i=0}^{s-1} x_i\gamma_i$, we
obtain the expansion map
\[
\Gamma:\mathbb{F}_{p^s}\to \mathbb{F}_p^s,\;
x\mapsto (x_0,\ldots,x_{s-1}),
\]
which extends componentwise to vectors and matrices.
For a code $\mathcal C\subseteq \mathbb{F}_{p^s}^n$, its \emph{expanded code} is
$ 
\Gamma(\mathcal C)=\{\Gamma(c)\in \mathbb{F}_p^{sn}\mid c\in \mathcal C\},
$
which is an $[sn,sk]_p$ linear code. If $G$ is a generator matrix of
$\mathcal C$, then a generator matrix of $\Gamma(\mathcal C)$ is obtained by
expanding each entry of $G$ with respect to $\Gamma$.
If $G'=SGP$, then
$
\Gamma(G')=\Gamma(S)\Gamma(G)(\text{Id}_s\otimes P),
$
so permutation equivalence is preserved. For linear equivalence, however, the
diagonal part causes an obstruction: if $G'=SGDP$ with
$D=\mathrm{diag}(d_1,\ldots,d_n)$, then the induced action on the expanded code
is monomial only when all $d_i\in \mathbb{F}_p$. Hence expansion yields an LEP
instance over the base field only for LEP($U$), with $U \subseteq \mathbb{F}_p^\star$.

Another natural descent is given by the trace map
$
\text{Tr}_{\mathbb{F}_{p^s}/\mathbb{F}_p}(\alpha)=\sum_{i=0}^{s-1}\alpha^{p^i}.
$
For a code $\mathcal C\subseteq \mathbb{F}_{p^s}^n$, define its \emph{trace code} by
$
\text{Tr}(\mathcal C)=\{(\text{Tr}(c_1),\ldots,\text{Tr}(c_n))\mid (c_1,\ldots,c_n)\in \mathcal C\}
\subseteq \mathbb{F}_p^n.$
This is an $[n,\le sk]_p$ linear code; moreover, it is the dual of a subfield
subcode, i.e., $\mathcal{C} \cap \mathbb{F}_p^n$.
If $G'=SGP$, then the corresponding trace codes are again permutation
equivalent. However, for trace codes, the diagonal part of a monomial
equivalence does not behave well with respect to the trace, since the scalars
$d_i$ cannot in general be pulled through the trace coordinatewise. Thus, while
both constructions naturally preserve permutation equivalence, neither gives a
general reduction of LEP over $\mathbb{F}_{p^s}$ to LEP over $\mathbb{F}_p$.

\paragraph{Connections to Other Incidence Structures:}

Another natural idea is to exploit the many connections between coding theory and other finite structures, e.g. projective systems, or matroids. 

Linear codes admit a geometric interpretation via projective systems. 
Let $G\in \mathbb{F}_q^{k\times n}$ be a generator matrix of a non-degenerate $[n,k,d]_q$ code $\mathcal C$. 
Considering the one-dimensional subspaces generated by the columns of $G$ yields a multiset of points 
$ \mathcal M \subseteq \mathrm{PG}(k-1,q), $
called the \emph{projective system} associated to $\mathcal C$.
This construction gives a one-to-one correspondence between non-degenerate $[n,k,d]_q$ codes and projective $[n,k,d]_q$ systems. 
Moreover, column permutations and non-zero column scalings do not change the corresponding points in projective space and hence the projective system.
Therefore, linearly equivalent codes give rise to equivalent projective systems.

Another closely related structure is the representable matroid associated to a generator matrix $G\in \mathbb{F}_q^{k\times n}$. 
Let $E=\{1,\ldots,n\}$ and define the independent sets $I=\{S\subseteq E \mid G_S \text{ has full rank}\}.$
This yields a matroid $M(G)$ whose rank function is given by
$ 
r(S)=\dim(\langle G_S\rangle).
$
If two codes are linearly equivalent, then the linear dependency relations among the columns of their generator matrices are preserved (up to relabeling of the coordinates), i.e., they define  isomorphic matroids.
This was also noticed in \cite{bennett2025relating}.
However, many questions about the relationships of the two problems and the complexity of matroid isomorphism remain open.

\paragraph{Open Questions}

Finally, we pose some open problems which we deem fruitful for further research and which might be (partially) in reach using existing techniques.

\begin{itemize}
 
\item Find (efficient) algorithms to find isomorphisms between matroids. 
\item Improve the complexity of the known reductions by achieving 
$\operatorname{poly}(n, \log(q))$ bounds for all reductions presented.
\item Are there reductions from LEP to LIP over non-prime fields?
    \item 
    Introduce an intermediate problem similar to UCE and find more reductions for Figure \ref{fig:reductions} including LEP$(U)$.
    \item Follow \cite{Mathon79} to get similar reductions between different versions of code equivalence problems.
    \item Show that random codes have  automorphism group $\{\lambda \; \text{id} | \lambda \in \F_q^\star \}$ with high probability for $q>2$.

    \item Give a search to decision reduction for LEP.
    \item Find an invariant which is characterizing.
    \item Find linear equations to model that $M \in M_{n,q}$.
    \item Which information can be found on $\varphi \in M_{n,q}$ through side-channel attacks?
\end{itemize}

\bibliography{biblio}
\bibliographystyle{plain}

\end{document}